\documentclass{ptephy_v1}%%%%%% to generate preprint number with ptep logo

\preprintnumber{XXXX-XXXX} %%% %%% Insert preprint number here

\usepackage[normalem]{ulem}  % \sout{old text} for strikeout
\usepackage{bm,amssymb}
\usepackage{txfonts}
\usepackage{mydefs}
\renewcommand{\sout}{\bgroup \color{red} \ULdepth=-.5ex \ULset}
\begin{document}

\title{
Chiral Symmetry and hadron properties at finite temperature
-- A numerical experiment
}

%%%% To generate auto affiliation numbers please use \author{}\affil{} command

\author[1]{Teiji Kunihiro}
\affil[1]{Department of Physics, Kyoto University, Kyoto 606-8502, Japan}
\author[2]{Shin Muroya}
\affil[2]{Faculty of Comprehensive Management, Matsumoto University, Matsumoto 390-1295, Japan}
\author[3,4,5]{Atsushi Nakamura}
\affil[3]{School of Biomedicine, Far Eastern Federal University, 690950 Vladivostok, Russia}
\affil[4]{Research Center for Nuclear Physics (RCNP), Osaka University, Ibaraki, Osaka, 567-0047, Japan}
\affil[5]{Theoretical Research Division, Nishina Center, RIKEN, Wako 351-0198, Japan}
%\affil[6]{IMC, Hiroshima University, Higashi-Hiroshima 739-8521, Japan}
\author[6,7]{Chiho Nonaka}
\affil[6]{Kobayashi Maskawa Institute, Nagoya University, Nagoya 464-8602, Japan}
\affil[7]{Department of Physics, Nagoya University, Nagoya 464-8602, Japan} 
\author[8]{Motoo Sekiguchi}
\affil[8]{School of Science and Engineering, Kokushikan University, Tokyo 154-8515, Japan}
\author[8]{Hiroaki Wada}
\author[3,5]{Masayuki Wakayama}
%\author{Insert third author name here}
%\author[3]{Insert fourth author name here} %%% Use optional bracket [3] to change the respective address
%\affil{Insert third author address here}

%%% To include the collaborator name... Please use the command "\collaborator"
%%% For example: \collaborator{ATLAS Collaboration}

\begin{abstract}%
We study the hadron properties at finite temperature from measurement of the screening masses, 
using two-flavor full QCD  of the hybrid Monte Carlo (HMC) 
algorithm with the renormalization group improved Iwasaki gauge action and the clover improved Wilson 
quark action on a $16^3 \times 4$ lattice.
We explore rather heavy quark mass regions. 
Disconnected quark diagram is dropped.
We observe  
the tendency that the screening masses in all the channels degenerate, which is in accord with the effective restoration of 
U$_{\rm A}$(1) symmetry, and then eventually approach 2$\pi T$, i.e. the free quark value. 
In the low temperature region below pseudocritical temperature $T_c$, 
the screening masses in all the channels decrease.
We discuss the different features between these calculations and
the previous ones.
\end{abstract}

\subjectindex{xxxx, xxx}

\maketitle

\section{Introduction}
Hadrons are elementary excitations on top of the nonperturbative QCD (Quantum Chromodynamics) vacuum 
where (colored) free quarks and gluons do not exist in the asymptotic states. 
The properties are called color confinement.
Furthermore, the QCD vacuum is characterized 
by the dynamical breaking of (approximate) chiral symmetry (DBCS) and U$_{\rm A}$(1) anomaly.
The DBCS manifests itself as the appearance of the nearly massless pseudoscalar mesons with the 
peculiar coupling properties.  
On the other hand, the axial anomaly is, for instance, responsible for the large mass of
$\eta'(958)$ compared with the other low-lying eight pseudoscalar mesons 
and the small mixing angle with $\eta(550)$ in the flavor SU(3)$_{\rm f}$ basis.
Moreover the axial anomaly can account for the large mass difference of 
the pion and the a$_0$.  

A unique feature of the QCD vacuum is that it should undergo 
a phase transition to chiral-symmetric phase at high temperature ($T$) and/or high baryon chemical potential. 
The axial anomaly may also be effectively restored in such an extreme condition \cite{Pisarski:1983ms}.  
Thus we can expect that hadrons on top of the new ground state or thermal equilibrium state
may change their properties along with that of the QCD vacuum
\cite{Hatsuda:1984jm,Hatsuda:1985eb,Hatsuda:1985ey,Hatsuda:1986gu};
in turn, (colorless) hadronic excitations with a small width may exist in the low-energy  %regime 
region 
even in the quark-gluon plasma (QGP) phase as the soft modes of the chiral transition
\cite{Hatsuda:1984jm,Hatsuda:1985eb} and the hydrodynamic modes \cite{DeTar:1985kx}.

The restoration of chiral symmetry should lead to degeneracy of the hadron spectral functions 
in the channels of the chiral partners: in the chiral SU(2)$_{\rm L}\times$SU(2)$_{\rm R}$ symmetry, 
the chiral partner of the pion is the sigma meson (isoscalar-scalar) while that of the isovector-vector meson $\rho$ is 
the isovector-axial vector meson a$_1$. 
On the other hand, the pion and a$_0$ mesons are  the partner in terms of  U$_{\rm A}$(1) symmetry.
Conversely speaking, the rate of the degeneracy of the spectral functions in the  channels of the chiral
partners  can be used as 
a measure of the symmetry restoration.

The first calculation of the chiral partners in the scalar channels based on a chiral effective model
\cite{Hatsuda:1984jm,Hatsuda:1985eb,Hatsuda:1985ey,Hatsuda:1986gu} showed that the sigma meson becomes 
soft and tends to get degenerate
with the pion  near the pseudocritical temperature $T_{c}$ of the chiral phase transition. 
QCD sum-rule calculations \cite{Hatsuda:1992bv} showed that 
the masses of the $\rho$ and its chiral partner a$_1$  both decrease at finite $T$, although 
their possible degeneracy toward the critical temperature is obscure depending on the calculation method.   

In the present work, we explore the hadron properties at finite temperature using the lattice simulations
of QCD. 
We study the screening masses \cite{DeTar:1987ar,DeTar:1987xb} obtained 
from the current-current correlations in the spatial direction.
Indeed, some authors \cite{Bazavov:2014cta,Maezawa:2016pwo,Brandt:2016daq}
performed lattice-QCD simulations of the static screening masses at finite temperature
including the critical (crossover) temperature and tried to extract the 
strength of the U$_{\rm A}$(1) anomaly in the chiral limit as well as the behavior of the 
screening masses of the chiral partners in the scalar and vector channels:
in Ref.~\cite{Maezawa:2016pwo} where the highly improved staggered quark (HISQ) action is used 
as in Ref.~\cite{Bazavov:2014cta}, 
it is shown that the screening masses of the positive parity mesons decrease whereas 
those of the negative parity mesons monotonically increase with temperature and thereby tend to get
degenerate; they eventually approach  $2\pi T$ above
the critical temperature. This behavior is consistent with the chiral degeneracy and shows a
possible scenario of the way how the chiral symmetry is restored in the vector and axial-vector channels. 
It is, however,  
%worth mentioning 
to be noted that this scenario is quite different from that suggested in 
\cite{Hatsuda:1992bv}; see also a recent analysis at finite density \cite{Gubler:2016djf}.
It is conceivable that the hadronic modes are modified by usual thermal effects as described 
by the hadron resonance gas model \cite{Bazavov:2012jq}; 
the intrinsic QCD dynamics 
becomes an essential ingredient for determining the properties of the hadronic modes just around 
the critical point.
Thus it should be helpful for identifying the underlying mechanism of spectral changes of
hadronic modes   
 to explore 
the thermal behavior of the screening masses at finite $T$ using  different quark actions 
with various quark masses.

We dare to make simulations of the screening masses using  rather heavy quark masses. 
In this way, the number of 
thermally excited hadrons are 
suppressed, and consequently,
they would hardly contribute to 
the thermodynamics of the system. 
We expect that such a simulation of the extreme set up far away from the chiral limit
 may provide us with 
the basic ingredients to realize the spectral changes of hadrons obtained 
in more complete lattice simulations. 

We find that below $T_c$,  not only positive-parity screening masses 
but negative-parity screening masses also   decrease in contrast to 
the findings given by the simulations respecting chiral symmetry \cite{Maezawa:2016pwo,Brandt:2016daq}.  
As for the effective restoration of U$_{\rm A}$(1) symmetry, for which the intrinsic dynamics of QCD is responsible, 
our simulation shows that it manifests itself 
as a drop of the positive-parity
screening masses and an increase of the corresponding 
 negative-parity screening masses 
above 
$T_c$  and degenerate at very high temperature in accord with the previous simulations with
different quark actions \cite{Maezawa:2016pwo,Brandt:2016daq}.  

The paper is organized as follows: in Sec.~\ref{sec:lattice}, we explain simulation parameters of our two-flavor 
full lattice QCD calculation. 
In Sec.~\ref{sec:results}, we show calculated results of screening masses of scalar, pseudoscalar, vector 
and axial-vector channels as a function of the gauge coupling $\beta$ and 
the ratio of temperature to the pseudocritical temperature $T/T_c$. 
We also discuss quark mass dependence of them. 
Section \ref{sec:summary} is devoted to summary and discussion. 

\section{Lattice simulations \label{sec:lattice}} 
We generate the gauge configurations in two-flavor full QCD  using the hybrid Monte Carlo (HMC) 
algorithm with the renormalization group improved Iwasaki gauge action and the clover improved Wilson quark action on 
a $16^3 \times 4$ lattice.  
Here we investigate hadron properties at finite temperature with  
relatively heavy quark masses, 
hoping to identify the basic ingredients for realizing the  temperature dependence of screening 
masses, using different gauge and fermion actions in lattice QCD. 
The pseudo-critical temperature $T_c$ in the present work is determined 
from the peak position of the susceptibility of the Polyakov loop, 
which is supposed to describe the confinement-deconfinement transition 
modified by the presence of quarks even if their masses are heavy. 
 We use the same combinations of the gauge couplings $\beta$  and hopping parameters $\kappa$ 
as those in Ref.~\cite{Ejiri:2009hq} in which they also evaluate the corresponding temperature.   
The coefficient $c_{SW}$ in the clover improved Wilson quark action is given by 
$c_{SW}=(1-0.8412 \beta^{-1})^{-3/4}$ \cite{Ejiri:2009hq}. 
The simulation parameters are listed in Table~\ref{tab:param}. 
In the case of $m_{\rm ps}/m_{\rm v}=0.75$ values of ratio $T/T_{c}$ do not contain errors, 
because we read off values of $\kappa$  and $\beta$ along the $m_{\rm ps}/m_{\rm v}=0.75$ line 
from Fig.~1 in Ref.~\cite{Ejiri:2009hq}. 
We update the first 1500 trajectories in the quenched QCD and switch to a simulation with the dynamical 
fermion. 
We discard the next 2000 trajectories of HMC and start to save the gauge configurations every ten 
trajectories. 
%%%%%%%%%%%%%%%%%%%%%%%%%%%%%%%%%%%%%%
\begin{table}[!h]
\caption{Simulation parameters for $m_{\rm ps}/m_{\rm v}=0.65, 0.75$ and 0.80 \cite{Ejiri:2009hq}. We 
produce 1000 gauge configurations for each parameter. }
\label{tab:param}
\begin{center}
\begin{tabular}{ccc|ccc}
\hline 
\hline 
\multicolumn{3}{c|}{$m_{\rm ps}/m_{\rm v}=0.65$} & \multicolumn{3}{|c}{$m_{\rm ps}/m_{\rm v}=0.80$} \\
\hline 
$\beta$  & $\kappa$ & $T/T_{c}$ & $\beta$  & $\kappa$ & $T/T_{c}$ \\
\hline 
1.90      & 0.141849  & 1.32(5)  &   1.60 &  0.143749     & 0.80(4)   \\
2.00      &  0.139411 & 1.67(6)  &  1.70  &  0.142871      & 0.84(4)\\
\cline{1-3} 
\multicolumn{3}{c|}{$m_{\rm ps}/m_{\rm v}=0.75$} & \multicolumn{1}{c}{1.80} &   \multicolumn{1}{c}{0.141139} &  \multicolumn{1}{c}{0.93(5)} \\
\cline{1-3}
$\beta$ & $\kappa$ & $T/T_{c}$ &  1.85  &   0.140070    & 0.99(5)   \\
\cline{1-3}
1.55      & 0.146479 & 0.80  &  1.90  &  0.138817     &  1.08(5)  \\
1.96      & 0.138732 & 1.35 & 1.95   &     0.137716   & 1.20(6)\\
2.06      &  0.137254 & 1.70 & 2.00   &     0.136931   & 1.35(7) \\
\cline{1-3} 
             &             &  & 2.10   &     0.135860   & 1.69(8) \\
             &             &  & 2.20   &     0.135010   & 2.07(10) \\
              &             &  & 2.30   &     0.134194   & 2.51(13) \\
              &             &  & 2.40   &     0.133395   & 3.01(15) \\
 \hline 
\end{tabular}
\end{center}
\end{table}
%%%%%%%%%%%%%%%%%%%%%%%%%%%%%%%%%%%%%%

Assuming that mesons are composed of two quarks, we employ simple point-like sources and sinks  
for the construction of meson operators, 
\begin{equation}
M(x)= \sum^{3}_{c=1} \sum^{4}_{\alpha,\beta=1} \bar{q}^c_\alpha (x) \Gamma_{\alpha  \beta}  q^c_\beta(x),  
\label{Eq:meson}
\end{equation}
where $q(x)$ is the Dirac operator for the $u/d$ quark and 
$\Gamma_{\alpha \beta}$'s {stand for 
$I$, $\gamma_5$, $\gamma_\mu$ and $\gamma_\mu \gamma_5$ for the scalar, the pseudoscalar, the vector 
and the axial-vector channels.
The indices $c$ and $\alpha$ are the color and Dirac-spinor indices, respectively. 
Here we measure spatial correlation functions of mesons in the $z$ direction. 
The spatial correlation function of the scalar channel is composed of connected and disconnected diagrams. 
{In order to} calculate the disconnected diagram, we employ the $Z_2$ noise method and subtract the vacuum expectation 
value of them \cite{Kunihiro:2003yj}.

\section{Calculated results \label{sec:results}}
In Fig.~\ref{fig:mass-b-T} we show screening masses of the  pseudoscalar, the scalar,  the vector  and the axial-vector 
channels as a function of the gauge coupling $\beta$ and $T/T_{c}$. The statistical errors are estimated by the jackknife 
method with the bin size of 10. 
We extract the screening masses from exponential damping of the spatial correlation functions in $3<z<15$. 
In the scalar channel, we obtain the connected part of the spatial correlation with small errors; 
however, we confront severe noise in the calculation of the disconnected part even on the 1000 gauge configurations.  
Therefore, in this paper, we evaluate the screening masses of the scalar channel  by using only the connected part.  
In the two-flavor QCD the connected part of the scalar channel 
can be regarded as the a$_0$ meson.
%%%%%%%%%%%%%%%%%%%%%%%%%%%%%%%%%%%%%%
\begin{figure}[!h]
  \begin{minipage}{0.5\hsize}
  \centering
  \includegraphics[width=6.2cm]{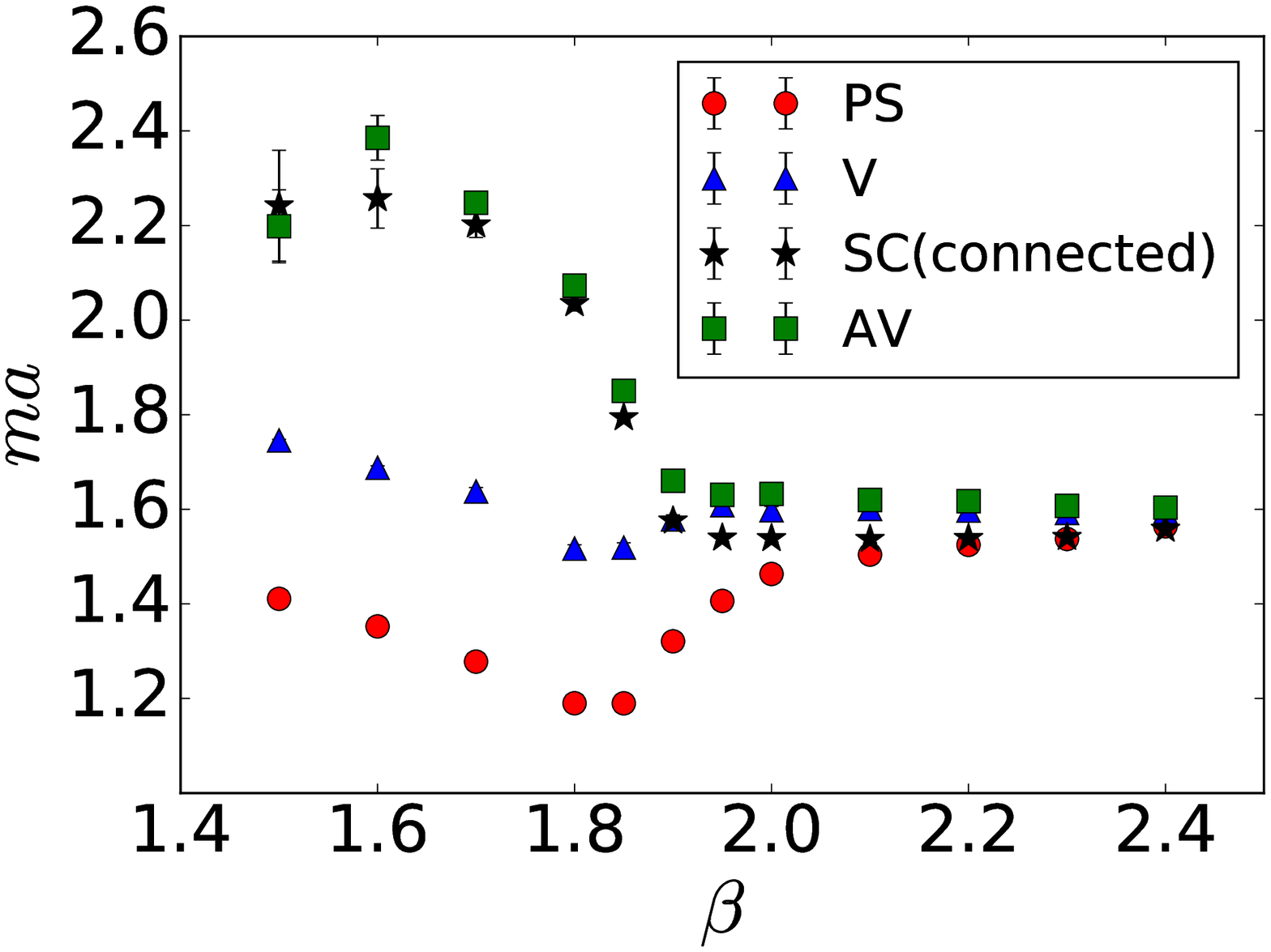}
  \end{minipage}
  \begin{minipage}{0.5\hsize}
  \centering
  \includegraphics[width=6.2cm]{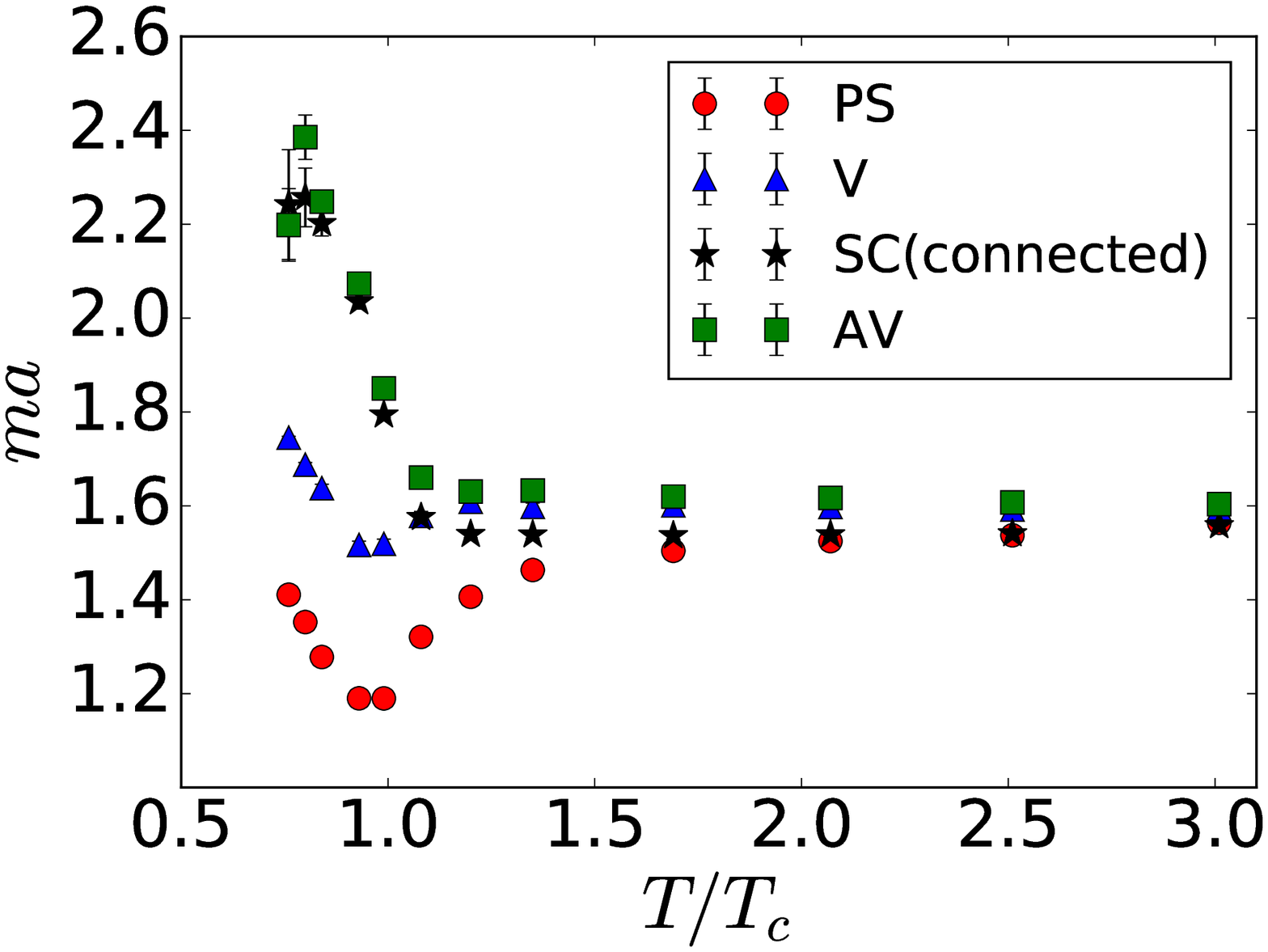}
  \end{minipage} 
  \vspace{1.5cm}
  \caption{The screening masses as a function of the gauge coupling $\beta$ (left) and 
  the ratio of temperature to the pseudocritical temperature $T/T_{c}$ (right), in the case of $m_{\rm ps}/m_{\rm v}=0.80$.  
 The red solid circles, the blue solid triangles, the black solid stars and the green solid squares 
 denote the pseudoscalar (PS), the vector (V), the connected part of the scalar (SC),  
 the axial-vector (AV) channels, respectively. 
 \label{fig:mass-b-T} }
\end{figure}
%%%%%%%%%%%%%%%%%%%%%%%%%%%%%%%%%%%%%%
In the low temperature region below the pseudocritical temperature ($T/T_{c} < 1.0$), 
the screening masses 
in all the channels decrease with the gauge coupling $\beta$ or the ratio $T/T_{c}$, 
except for the positive-parity hadrons at quite low temperature region. 
In particular, the screening masses of the scalar meson 
are heavier than those 
of the vector meson and almost the same as the ones 
of the axial-vector meson.  
This result 
is consistent with the fact that the a$_0$ meson mass is heavier than the $\rho$ meson mass. 
If 
we could include the disconnected diagram as well as the connected diagram in evaluation of screening masses 
in  
the scalar channel, we would obtain lower screening masses which are almost the same as those in
 the vector  
channel \cite{Kunihiro:2003yj}. 
It is noteworthy that a simulation
at almost physical quark mass on the 
gauge configurations generated in the (2+1)-flavor QCD with the HISQ action \cite{Bazavov:2014cta, Maezawa:2016pwo}
 shows an increase of the screening 
masses in the  negative parity channels, i.e. the pseudoscalar and vector channels
along with increasing temperature in the low temperature region. 
Since the hadron resonances are not expected to contribute to 
the dynamics in our simulation with heavy quarks, the decrease %found \cite{Bazavov:2014cta, Maezawa:2016pwo}
might be mainly attributed to mutual interactions of hadron resonances but not to the possible
change of the QCD vacuum.

Around the pseudocritical temperature $T_{c}$, the screening masses of negative parity mesons 
take a minimum value and above it they turn to increase with temperature. 
On the other hand, the screening masses of positive parity mesons, the scalar and axial-vector channels 
keep decreasing with temperature. Then around $\beta=1.95$ ($T/T_{c}=1.20$), 
the screening masses of negative parity mesons and 
positive parity mesons approach  each other and degenerate at $\beta=2.3$ ($T/T_{c}=2.5$), i.e.  
much the same behavior as those found in Refs.~\cite{Bazavov:2014cta, Maezawa:2016pwo}.
The degeneracy of the screening masses between the vector and the axial-vector channels suggests that 
the chiral symmetry is realized in this high temperature  region 
although the symmetry is badly broken with heavy quark mass in the present
simulation.
As for the scalar and pseudoscalar channels, it is noteworthy that 
% a drop of the screening mass in the scalar channel 
 the degeneracy in the scalar and pseudoscalar channels
is realized  
in spite of lack of the disconnected diagram in the 
scalar channel.  
Again this behavior implies that the 
 chiral symmetry gets realized in the high temperature region in spite of the large current 
quark mass.
 Furthermore, the U$_{\rm A}$(1) anomaly is a key issue to understand relations between 
the pseudoscalar and scalar channels at finite temperature. 
In fact, in Ref.~\cite{Brandt:2016daq} they tried to extract 
the strength of the U$_{\rm A}$(1) anomaly 
from the mass gap of screening masses between the pseudoscalar and scalar channels in the chiral limit. 
However, because our current calculation is performed with relatively heavy quark, 
small lattice size and neglecting 
the disconnected diagram in the scalar channel we do not go further in the discussion on it.  

At the highest temperature $\beta=2.4$ ($T/T_{c}=3.0$) we find 
that 
the screening masses in all the channels tend to get degenerate and approach $2 \pi T$. 

In Fig.~\ref{fig:mass-T} we show the screening masses of the pseudoscalar, the vector, 
the scalar from the connected diagram and the axial-vector channels as a function of $T/T_{c}$ 
in the case of $m_{\rm ps}/m_{\rm v}=0.65$, 0.75 and 0.80. 
For $m_{\rm ps}/m_{\rm v}=0.75$ ($m_{\rm ps}/m_{\rm v}=0.65$), we evaluate the screening masses in each channel at 
$T/T_c=0.80, 1.35$ and 1.70 ($T/T_c=1.32$ and  1.67). 
For the axial-vector channel, due to the large errors in spacial correlation functions 
we fail to evaluate the screening mass at $T/T_c=0.80$ in the case of $m_{\rm ps}/m_{\rm v}=0.75$.  
We observe the tendency that the screening masses of all channels decrease with the ratio $m_{\rm ps}/m_{\rm v}$. 
In particular, we find a distinct drop of the screening  mass in the pseudoscalar channel 
as the ratio $m_{\rm ps}/m_{\rm v}$ becomes smaller. 
On the other hand, in the vector channel we observe relatively weak dependence on $m_{\rm ps}/m_{\rm v}$. 
%%%%%%%%%%%%%%%%%%%%%%%%%%%%%%%%%%%%%%
\begin{figure}[!h]
\begin{center}
 \includegraphics[width=0.8\hsize]{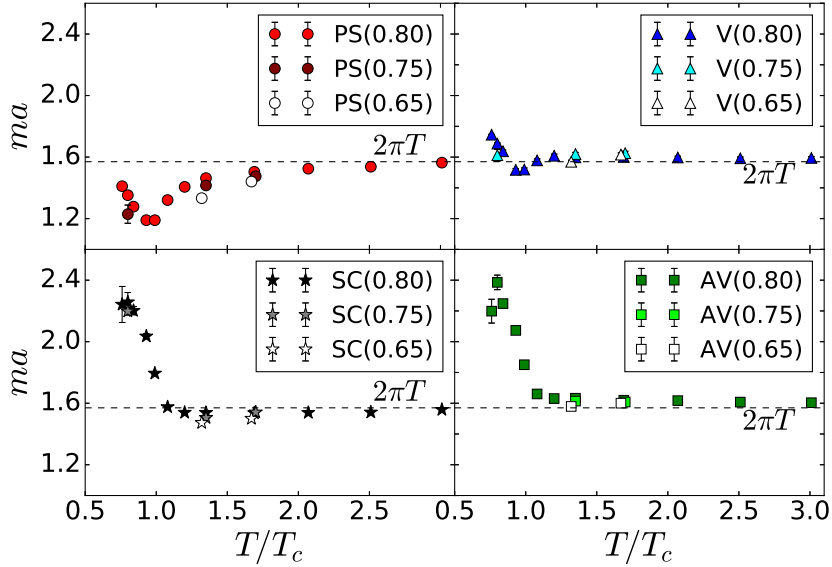}
 \end{center}
 \vspace{1cm}
  \caption{The screening masses of the pseudoscalar (PS) (top left), 
  the vector (V) (top right), the connected part of the scalar (SC) (bottom left) and 
the axial-vector (AV) (bottom right) channels as a function of $T/T_{c}$ in the case 
of $m_{\rm ps}/m_{\rm v}=0.65$, 0.75 and 0.80. 
The dotted lines stand for $2\pi T$.
\label{fig:mass-T} }
\end{figure}
%%%%%%%%%%%%%%%%%%%%%%%%%%%%%%%%%%%%%%

Figure \ref{fig:mass-qm} shows the screening masses 
in the  pseudoscalar,  the vector and the scalar channels 
from the connected diagram 
and the axial-vector channels as a function of the ratio $m_{\rm ps}/m_{\rm v}$, 
at $T/T_{c}=1.69$ (top), $T/T_{c}=1.35$ (bottom left) and $T/T_{c}=0.80 $ (bottom right). 
At first, the screening masses in all the channels become smaller
with  lighter quark masses, but the decreasing 
rate of them is different in each channel. 
In the case of $T/T_{c}=1.69$,  at the $m_{\rm ps}/m_{\rm v}=0.80$ the  
complete degeneracy 
between the vector and the axial-vector channels is found; 
however, it tends to resolve as the quark mass becomes smaller.  
We can see weaker trend to degeneracy between the pseudoscalar and 
the connected part of the scalar channels 
at $m_{\rm ps}/m_{\rm v}=0.80$, but it also disappears at lighter quark masses. 
At $T/T_{c}=1.35$ the difference of screening masses between the pseudoscalar and the vector channels is larger than that at 
$T/T_{c}=1.69$. 
To obtain a rough idea of the value of screening mass of each channel at the physical quark mass, 
we draw the linear line which is obtained by fitting screening masses at $m_{\rm ps}/m_{\rm v}=0.65$, 0.75 and 0.80. 
The thick solid line represents the value of the physical quark mass. 
Because our quark masses are far from the physical value, the extrapolation should not be taken to be serious. 
However, it suggests that the clear degeneracy observed in our calculation 
might be  a 
behavior 
specific to 
heavy quarks.
%%%%%%%%%%%%%%%%%%%%%%%%%%%%%%%%%%%%%%
\begin{figure}[!h]
  \centering
  \includegraphics[width=6.0cm]{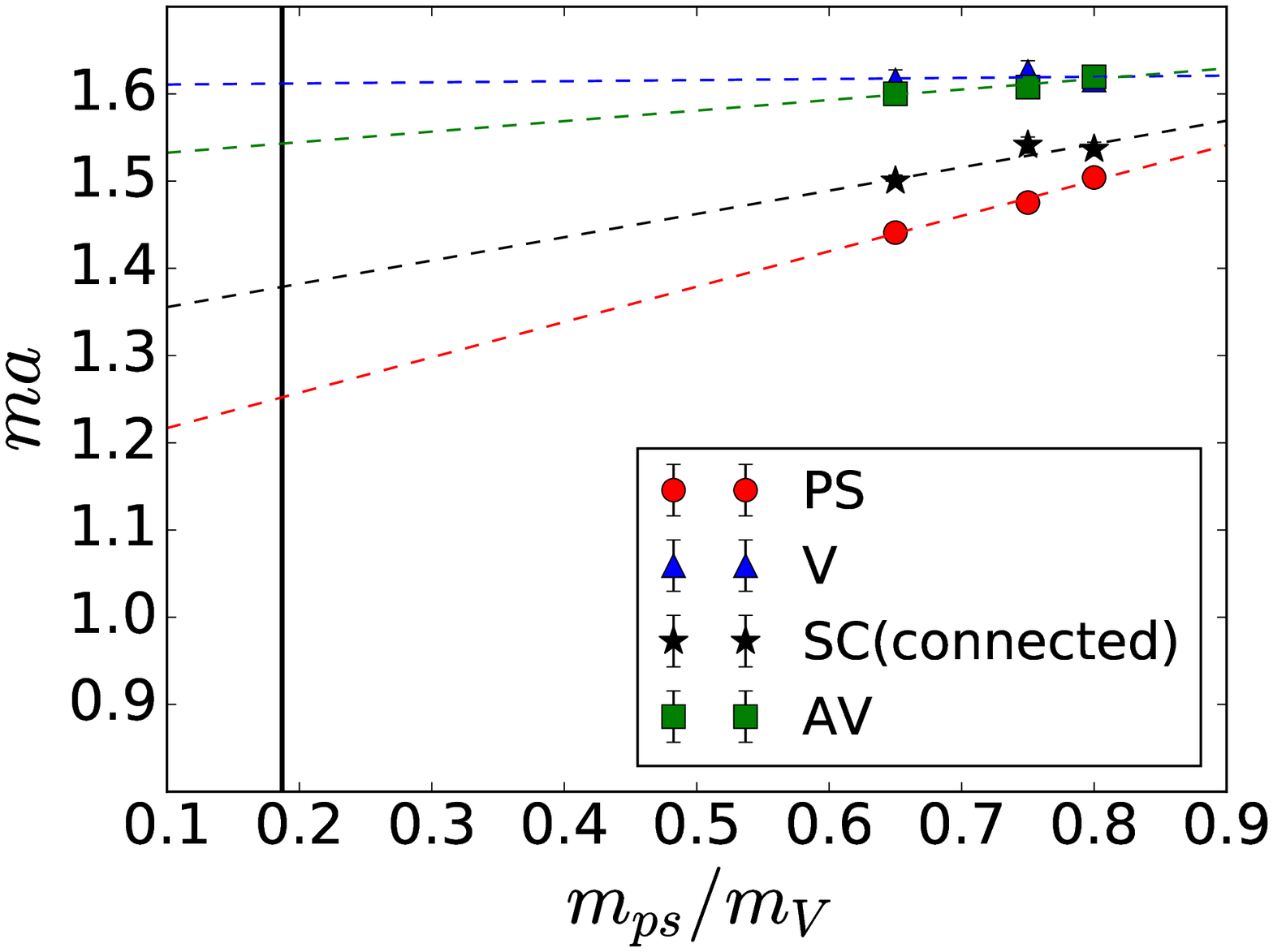}\\
 \begin{minipage}{0.48\hsize}
  \centering
  \includegraphics[width=6.0cm]{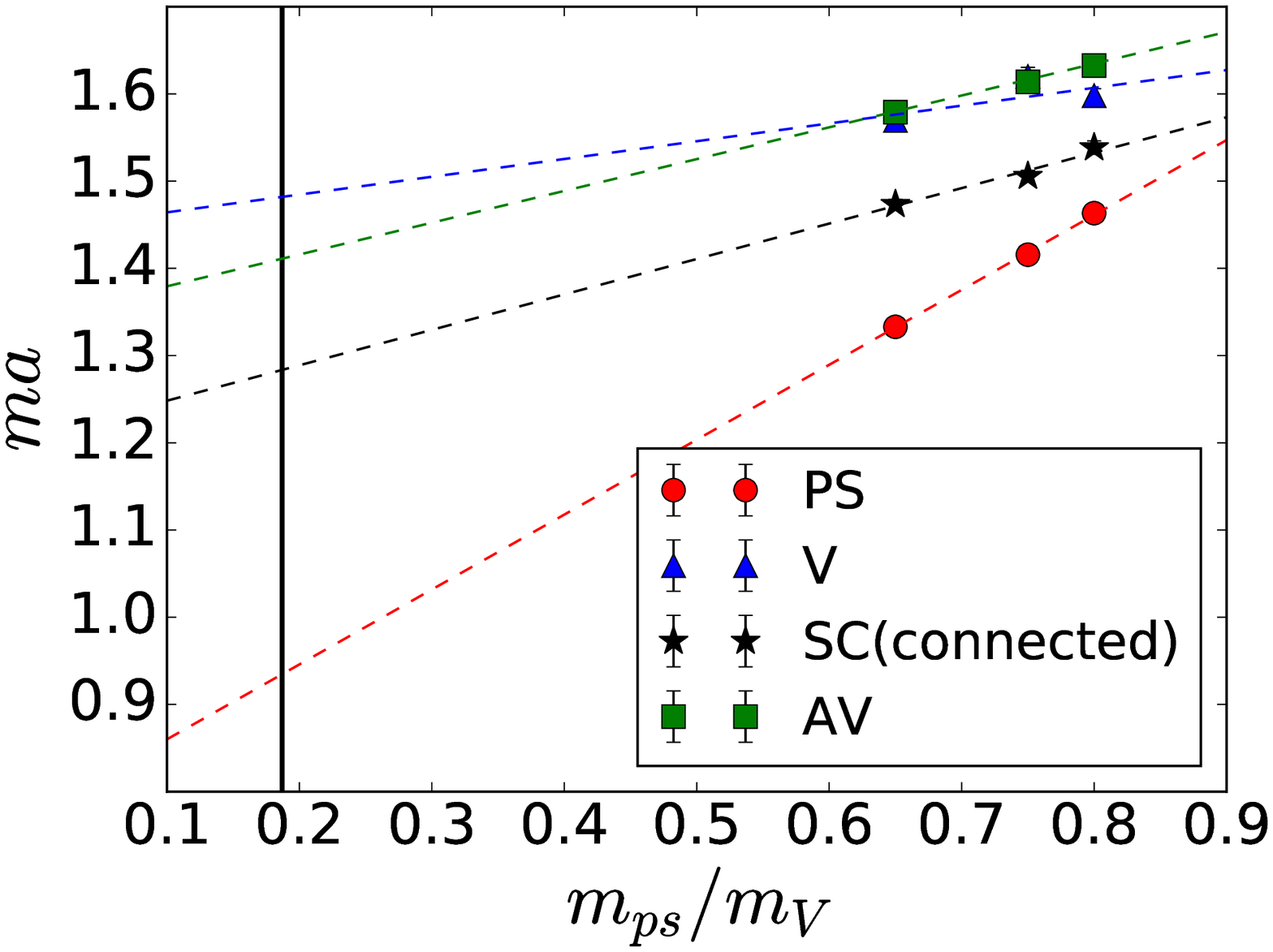}
  \end{minipage} 
    \begin{minipage}{0.48\hsize}
  \centering
   \includegraphics[width=6.0cm]{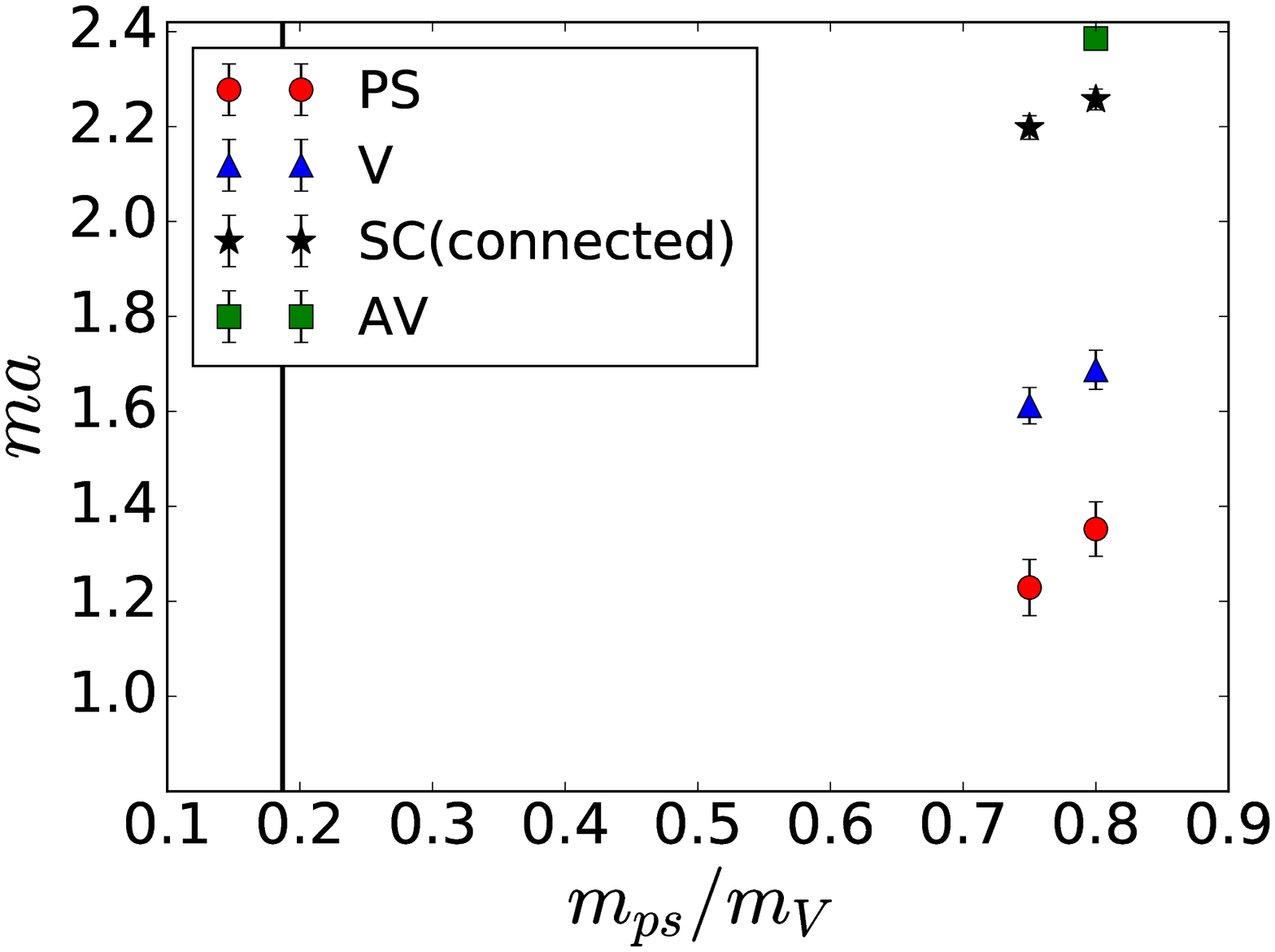}
  \end{minipage} 
  \vspace{1.5cm}
  \caption{The screening masses of the pseudoscalar (PS), the vector (V), the connected part of 
  the scalar (SC) and the axial-vector (AV) channels as a function of 
 the ratio $m_{\rm ps}/m_{\rm v}$ in the case of $T/T_{c}=1.69$ (top), $T/T_{c}=1.35$ (bottom left) and $T/T_{c}=0.80 $ (bottom right), respectively. 
\label{fig:mass-qm} }
\end{figure}
%%%%%%%%%%%%%%%%%%%%%%%%%%%%%%%%%%%%%%

\section{Summary \label{sec:summary} and discussion}
We have explored the hadron properties at finite temperature using two-flavor full QCD  of the hybrid Monte Carlo (HMC) 
algorithm with the renormalization group improved Iwasaki gauge action and the clover improved Wilson 
quark action on a $16^3 \times 4$ lattice.
In spite of the limited calculations with heavy quark mass, $m_{\rm ps}/m_{\rm v}=0.80$, 
we have observed that at $T/T_c=2.5$ the clear degeneracy between the screening masses of 
the negative parity mesons and the positive parity mesons. 
At the highest temperature $\beta=2.4$ ($T/T_{c}=3.0$) we find the tendency that 
the screening masses in all the channels degenerate in accord with the effective restoration of U$_{\rm A}$(1) symmetry, 
and then eventually approach $2\pi T$, namely, the free-quark value.

In the low temperature region below pseudocritical temperature $T_c$, the screening masses 
in all the channels 
decrease along with temperature, which is a different 
behavior from that  found in the previous calculation 
obtained at almost physical quark mass
on the gauge configurations generated  in the (2+1)-flavor QCD with the HISQ action
 \cite{Bazavov:2014cta, Maezawa:2016pwo}.
Our finding of the behavior of the screening masses may be a 
characteristic feature in the  heavy quark sector 
because hadron gas resonance dynamics would not play any significant role in contrast to
the simulations respecting chiral symmetry \cite{Bazavov:2014cta, Maezawa:2016pwo}.
Furthermore, we make a comment on a fermion lattice action. Here we utilized the clover improved Wilson 
quark action in which the chiral symmetry is explicitly broken. 
To elucidate physics related with the chiral symmetry, chiral fermion actions such as domain wall fermions and 
overlap fermions would be more suitable.

We have also investigated the quark mass dependence of behavior of the screening masses at finite temperature. 
The degeneracy which we find at high temperature above $T_c$ between 
screening masses of the negative parity mesons and the positive parity mesons is resolved at lighter quark masses. 
To reach  a conclusive result, we need to check the detailed behavior of screening masses of each channel 
such as lattice size and quark mass dependence of them.

\section*{Acknowledgment}
The work of C.N. is supported by the JSPS Grant-in-Aid for Scientific Research (S) No. 26220707 and 
the JSPS Grant-in-Aid for Scientific Research (C) No. 17K05438. 
The work of T.K. and A.N. is partially supported by the JSPS Grant-in-Aid 
for Scientific Research (B) No. 15H03663. 
The work of S. M. is partially supported by the Grant-in-Aid of Matsumoto University for the Academic Research.  The simulation was performed on an NEC SX-9 and SX-ACE supercomputers at RCNP, Osaka University, and was conducted with the Fujitsu PRIMEHPC FX10 System (Oakleaf-FX, Oakbridge-FX) in the Information Technology Center, the University of Tokyo.


\bibliographystyle{ptephy}
\bibliography{mybibfile}
%
% once the .bbl file has been generated then place the text in your article.

%\begin{thebibliography}{9}

%\bibitem{1}
%J. P. Blaizot, and E. Iancu, Phys. Rep. {\bf 359}, 355 (2002).

%\bibitem{2}
%M.~Gyulassy, and L.~McLerran, Nucl.\ Phys.\  A {\bf 750}, 30 (2005).

%\bibitem{3}
%U. W. Heinz, and P. F. Kolb, Nucl. Phys. {\bf A702}, 269 (2002).

%\bibitem{4}
%T.~Hirano, U.~W.~Heinz, D.~Kharzeev, R.~Lacey, and Y.~Nara,
%Phys.\ Lett.\  B {\bf 636}, 299 (2006).

%\bibitem{5}
%R. Baier, A. H. Nueller, D. Schiff, and D. T. Son, Phys. Lett. B {\bf 502}, 51 (2001).

%\end{thebibliography}

%\appendix

%\section{Appendix head}

\end{document}